\documentclass[
letter,
reprint,
a4paper,
aps,prl,
amsmath,
]{revtex4-1}

\usepackage{mathtools,amssymb}
\usepackage{booktabs}
\usepackage{siunitx}
\usepackage[]{hyperref}
\usepackage{multirow}
\usepackage{bbold} 
\usepackage{float}

\newcommand{\Uni}[2]{\ensuremath{{#1} \mathrm{\:{#2}} }}

\newcommand{\pop}{\ensuremath{ \wp } }

\newcommand{\ket}[1]{\ensuremath{\left| {#1} \right>} }
\newcommand{\bra}[1]{\ensuremath{\left< {#1} \right|} }
\newcommand{\expect}[1]{\ensuremath{\langle #1 \rangle}}
\newcommand{\ketbra}[2]{\ensuremath{| {#1} \rangle \langle {#2} |}}

\newcommand{\braket}[2]{\ensuremath{\left< \left. {#1} \right| {#2} \right>}}

\newcommand{\inoket}{\ket{\psi_{\text{in}}}}

\newcommand{\genosc}{\ensuremath{\ket{\psi}} }

\newcommand{\bragenosc}{\ensuremath{\bra{\psi}} }
\newcommand{\oscin}{\ensuremath{ \ket{\psi_{\textrm{in}} } }}
\newcommand{\oscinbra}{\ensuremath{ \bra{\psi_{\textrm{in}} } }}
\newcommand{\oscout}{\ensuremath{\ket{\psi_{\pm 1}} }}

\newcommand{\create}{\ensuremath{{\,\hat{a}^{\dagger}}} }
\newcommand{\destroy}{\ensuremath{\,\hat{a}} }

\newcommand{\nbar}{\ensuremath{\, \langle n \rangle} }

\newcommand{\Dis}[1]{\hat{\mathcal{D}}(#1)}
\newcommand{\Sq}[1]{\hat{\mathcal{S}}(#1)}

\newcommand{\wig}{\ensuremath{\mathcal{W}} }
\newcommand{\charfun}{\ensuremath{\chi} }
\newcommand{\qfun}{\ensuremath{\mathcal{Q}} }

\newcommand{\Parity}{\ensuremath{\hat{\mathcal{P}}} }

\newcommand{\intd}{\ket{\downarrow}}
\newcommand{\intup}{\ket{\uparrow} }

\newcommand{\sigx}{\hat{X}}
\newcommand{\sigy}{\hat{Y}}
\newcommand{\sigz}{\hat{Z}}

\newcommand{\be}{\begin{eqnarray}}
\newcommand{\ee}{\end{eqnarray}}

\newcommand{\caf}{\ensuremath{^{40}{\textrm{Ca}}^{+}\, }}
\newcommand{\bef}{\ensuremath{{^9}{\textrm{Be}}^{+} \,}}

\usepackage[acronym]{glossaries}
\usepackage{etoolbox}
\newtoggle{arXiv}
\toggletrue{arXiv}

\let\oldcite\cite
\renewcommand{\cite}[1]{\mbox{\oldcite{#1}}}

\usepackage{tikz}
\usetikzlibrary{arrows,calc,matrix,fit,positioning}
\usetikzlibrary{decorations.pathreplacing}
\usetikzlibrary{decorations.pathmorphing}
\usetikzlibrary{decorations.markings}
\usetikzlibrary{shapes.geometric}
\usetikzlibrary{decorations.pathreplacing}
\usetikzlibrary{calc}
\usepackage{quantumcircuits}

\newcommand{\myarrow}[1][0]{%
  \mathrel{%
    \text{$
     \begin{tikzpicture}[baseline = -0.5ex]
    \end{tikzpicture}
    $}%
  }%
}%
\usepackage{quantumcircuits}

\tikzset{
	circuit/.style = {
		row sep=1ex,
		column sep=2em,
		nodes = {anchor=center},
	},
	gate/.style = {minimum width=7mm,minimum height=7mm,draw},
	Bgate/.style = {minimum width=10mm,minimum height=7mm},
	ctrl/.style = {circle,fill=black,inner sep=0.7mm},
	NOT/.style = {pluscircle,draw},
	meas/.style = {minimum width=7mm,minimum height=7mm,measurement,draw},
	state/.style = {},
	quantum bit/.style = {},
	classical bit/.style = {double=white,draw=black,double distance=0.6mm,line width=0.3mm},
}

\tikzset{	
pics/ASym2/.style args = {#1,#2,#3,#4,#5,#6}{
    code = {
			\pgfmathsetmacro\second{#1/2}%
				\pgfmathsetmacro\heig{#2/2}%
				\node[font=\tiny] (cen) at (\second,\heig){$\pm$};
        \node (le) at ($ (\second,\heig) + (-0.35,0) $){$\mathcal{D}($};
				\node[](h) at  ($ (le) + (-0.05,0.15) $) {$\hat{}$};
				\node (r) at ($ (\second,\heig) + (+0.55,0) $){$\alpha/2)$};			
				\draw ($ (\second,\heig) + (-0.12,-0.12) $) rectangle ($ (\second,\heig) + (0.12,0.12) $);
				\node[draw=black,minimum width=7mm,minimum height=7mm,inner sep=0pt, fit=(le.north west) (r.south east)] (DisB) {};
				
				\node[diamond,fill=black,minimum size=3mm , inner sep =0 pt](diam) at ($(\second,\heig) + (0,-#3)$) {};
				\draw (cen) -- (diam);
				
				\node[](inosc) at ($(\second,\heig) + (-#4,0)$) {$\oscin$};
				\node[](inqubit) at ($(\second,\heig) + (-#4,-#3)$) {$\intup$};
				\draw (inosc) -- (DisB);
				
				\node[measurement,minimum width=7mm,minimum height=7mm,inner sep=0pt,draw](mes) at ($(\second,\heig) + (#5,-#3)+ (#6,0)$) {};
				\node[right] at (mes.east) {$\expect{\sigz}$};
				
				\node[minimum width=7mm,minimum height=7mm,draw] (h2) at ($(\second,\heig) + (-#5,-#3)+ (#2,0)$) {$\hat{R}(\theta,\phi=0)$};
				\draw (inqubit) -- (h2) --(diam)--(mes);
				\node[](outosc) at ($(\second,\heig) + (#6,0)+ (#5,0)$) {$\oscout$};
				\draw (DisB) -- (outosc);
				
    }
  }
}

\begin{document}
\title{Direct characteristic--function tomography of quantum states of the trapped-ion motional oscillator}

\author{C. Fl{\"u}hmann and J.~P.~Home }
\email[Corresponding author, Email: ]{christaf@phys.ethz.ch}

\affiliation{Institute for Quantum Electronics, ETH Z\"urich, Otto-Stern-Weg 1, 8093 Z\"urich, Switzerland}

\date{\today}

\begin{abstract}
We implement direct readout of the symmetric characteristic function of quantum states of the motional oscillation of a trapped calcium ion. Using suitably chosen internal electronic state-dependent displacements based on bi-chromatic laser fields we map the expectation value of the real or imaginary part of the displacement operator to the internal states, which are subsequently read out using fluorescence detection. Combining the two readout results provides full information about the symmetric characteristic function. We characterize the performance of the technique by applying it to a range of archetypal quantum oscillator states, including displaced and squeezed Gaussian states as well as two and three component superpositions of displaced squeezed states which have applications in continuous variable quantum computing. For each, we discuss relevant features of the characteristic function and Wigner phase-space quasi-probability distribution. The direct reconstruction of these highly non-classical oscillator states using a reduced number of measurements is an essential tool for understanding and optimizing the control of these systems for quantum sensing and quantum information applications.
\end{abstract}


\maketitle

Quantum state reconstruction is an important element enabling diagnosis and improvement of quantum control. As larger states come under experimental control the number of measurements required to perform state reconstruction becomes crucial~\cite{16Chao}. Here, significant gains can be found by choosing the appropriate basis in which to make measurements~\cite{16Kienzler}. Bosonic systems such as mechanical harmonic oscillators and electromagnetic field modes play a prominent role across quantum information~\cite{01Gottesman,16Ofek}, quantum sensing~\cite{Aasi2013,Zhang18,Meyer01,18McCormick} and fundamental studies~\cite{15Arora,15Asadian2}. These have been prepared in a wide range of quantum states, including Fock, squeezed and displaced states~\cite{96Meekhof} and superpositions of all of these~\cite{96Monroe,Hofheinz2009}. Particular focus has been placed on states involving superpositions of displaced states, for which the archetypal example is the ``Schr{\"o}dinger's cat'' superposition of two displaced coherent states~\cite{Schrodinger1935}. Studies include evolution under decoherence channels ~\cite{00Turchette2, 09Wang} as well as storage and manipulation of information in error-correction codes~\cite{16Ofek,18Fluhmann2}.\par

Prior work on oscillator state tomography include techniques based on homodyne measurements~\cite{09Lvovsky} and methods based on extraction of Fock state occupations, parity and ground state occupation following displacements applied to the analyzed states~\cite{96Leibfried,Bertet02,97Lutterbach,13Vlastakis,Guerlin2007,Zhang18}.
These results are then processed to reconstruct states in the Fock state basis or to extract phase space quasi-probability distributions, such as the Wigner and Husimi $\qfun$-function. When extracting quasi-probabilities often a large amount of excess data is collected, such as extracting many Fock state occupations which are then reduced to a single parity value~\cite{96Leibfried,16Kienzler}, which makes this technique relatively expensive in terms of the number of measurements. Direct parity measurements have been used extensively for measuring the Wigner function in the context of electromagnetic field modes~\cite{Bertet02,13Vlastakis}, but for trapped-ions this remains relatively challenging~\cite{17Matsukevich,17Kim}. Measurement of ground-state population is restricted to extracting the Husumi-$\qfun$ function, but this is not well suited to analysis of cat-like states since quantum interference effects are exponentially suppressed with the separation of the displaced wavepackets~\cite{Supp}.\par
A complete description of the quantum state is also given by the characteristic function of the quasi-probability distributions ~\cite{02Barnett}. The symmetric characteristic function is defined as:
\be
\charfun(\beta)=\expect{\Dis{\beta}}
\ee
with $\Dis{\beta}=e^{\beta\hat a^{\dagger} -\beta^*\hat a}$ being a shift operator by the complex amount $\beta$~\cite{05Schleich}. Here $\hat{a}$ is the usual harmonic oscillator destruction operator and $\expect{\cdot}=\oscinbra \cdot \oscin$ denotes the expectation value evaluated on the analyzed state $\oscin$. The quasi-probability distributions can be obtained from the characteristic function using a two-dimensional Fourier transform of $\charfun(\beta)$~\cite{02Barnett}:
\begin{align}
&\wig_l(\gamma)=\frac{1}{\pi^2}\int \charfun(\beta) e^{l|\beta|^2/2} e^{\gamma \beta^*-\gamma^*\beta}d^2 \beta
\label{eq:WigFT}
\end{align}
For $l=0$ this gives the Wigner function, for $l=-1$ the Husimi-$\qfun$ function and for $l=1$ the Glauber-Sudarshan P-representation. These representations have different properties. The P-representation for example can become singular, while the Wigner and $\qfun$ function are both bounded. Thus only the latter two are commonly used in experiments~\cite{16Ofek,Wang1087,09Lvovsky,Bertet02,13Vlastakis,17Matsukevich,17Kim,96Leibfried,16Kienzler,09Wang}.
Methods to directly reconstruct $\charfun(\beta)$ have been proposed as early as 1995~\cite{Wallentowitz95,00Zheng} and have been used for 1D reconstruction of wavefunctions~\cite{12Casanova,11Gerritsma,10Gerritsma,10Zahringer} as well as 2D reconstruction of Fock and thermal states~\cite{15Johnson}.\par

In this Letter we perform direct reconstruction of the characteristic function of a trapped-ion motional oscillator state, and illustrate its use by applying it to a squeezed state, a displaced-squeezed state, a squeezed Sch{\"o}dinger's cat state \cite{16Kienzler, 15Lo} and a GKP codeword consisting of three superposed displaced-squeezed states \cite{01Gottesman, 18Fluhmann2}. All reconstructions are in close agreement with the independent calibrations of the measured states, but reveal small significant discrepancies in the experimentally set parameters, which could be used for future improvements in state control. The reconstruction method used is direct, requiring on average one readout setting to measure one characteristic function point for any input state. Thus we find a reduction in data taking time of more than a factor of 20 relative to methods we used previously~\cite{Supp}.
\begin{figure}[tb]
\begin{center}
\begin{tikzpicture}
\draw[] (0,-7) pic{ASym2={1.4,0.2,0.9,3.5,2,0.3}};
\end{tikzpicture}
\end{center}
\caption{Characteristic function readout circuit. A carrier rotation of fixed laser phase $\hat{R}(\theta,\phi=0)$ and an SDF pulse are applied to the internal state initialized to $\intup$ and the oscillator state under test $\oscin$. The diamond symbol in the circuit denotes control of the sign before the shift $\alpha/2$ in the $\sigx$ basis of the internal states. The final fluorescence readout statistics of the internal states follows the statistics $\expect{\sigz} = \cos(\theta)\textrm{Re} [\expect{\Dis{\beta} }] + \sin(\theta) \textrm{Im} [\expect{ \Dis{\beta}}]$ with $\charfun(\beta)=\expect{ \Dis{\beta}}$ the characteristic function.
}
\label{fig:CircuitChar}
\end{figure}
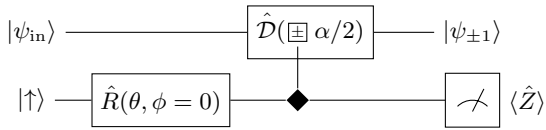

The oscillator used in the experiments is the axial motional mode of a single trapped \caf ion with a frequency of around $\omega_m \approx 2 \pi \times \Uni{1.9}{MHz}$. The motional mode is controlled and read out via the internal electronic levels $\intd \equiv \ket{^{2}{S}_{1/2}, m_j=1/2}$ and $\intup \equiv \ket{^{2}{D}_{5/2}, m_j=3/2}$. The quantum circuit used for reconstruction is given in figure~\ref{fig:CircuitChar}. At the beginning of each experiment the axial motional mode is initialized to a given oscillator state $\inoket$, which we aim to reconstruct. The initial electronic state is prepared to $\intup$. Then a resonant carrier rotation of angle $\theta$ and laser phase $\phi$ is applied to the internal states
$\hat{R}(\theta,\phi) = \cos(\theta/2)\mathbb{1} - i\sin(\theta/2)[\sin(\phi) \sigx + \cos(\phi) \sigy]$. Here $\sigx \equiv \ketbra{\uparrow}{\downarrow} + \ketbra{\downarrow}{\uparrow}$ and $\sigy \equiv -i\ketbra{\uparrow}{\downarrow} + i\ketbra{\downarrow}{\uparrow}$ are two Pauli matrices acting on the two internal states. This is followed by application of an internal state-dependent force (SDF) based on a bi-chromatic laser pulse realizing the operation $\Dis{\alpha(t)\sigx}$, where $\alpha(t)=\eta\Omega t e^{-i\Delta\phi/2}$~\cite{05Haljan}. Here $\eta \simeq 0.05$ denotes the Lamb-Dicke parameter~\cite{98Wineland2}, while $\Omega$ and $\Delta\phi$ are controlled via the total power and relative phases of the bi-chromatic laser fields. Thus the oscillator state is shifted with a direction dependent on the internal states being in either $\ket{+}$ or $\ket{-}$ defined as $\sigx\ket{\pm}=\pm\ket{\pm}$.
Finally the internal state is read out using resonant fluorescence~\cite{98Wineland2}. This circuit can be viewed as performing an indirect measurement of the oscillator via the internal states, which extracts modular position and momentum variables~\cite{18Fluhmann,10Popescu,15Asadian}. The internal state readout statistics follows
\begin{align}
\expect{\sigz} &=\expect{ e^{i\theta}\Dis{-\beta}+e^{-i\theta}\Dis{\beta} }/2 \\
&= \cos(\theta)\textrm{Re} [\expect{\Dis{\beta} }] + \sin(\theta) \textrm{Im} [\expect{ \Dis{\beta}}].\notag
\end{align}
The real part of the characteristic function can thus be obtained by choosing $\theta=0$, while the imaginary part is obtained from $\theta=\pi/2$. In the following we will use the short notation $\charfun(\beta)$ if the analyzed state $\oscin$ is unambiguous and $\charfun(\beta,\oscin)$ where the specific state is important. Note that any pair of angles differing by $\pi/2$ provide full information about the characteristic function $\charfun(\beta)$.\par

The characteristic function is complex valued and Hermitian $\chi(\beta)^*=\chi(-\beta)$ and thus any half of the complex space covered by $\beta$ is sufficient for a complete measurement. Therefore only a single fluorescence readout is required to obtain one characteristic function point. In our experiment this represents a reduction of two orders of magnitude of required readouts over previous work on Wigner function reconstruction~\cite{16Kienzler}. The number of sampled points can in principle be reduced further for suitable states by optimizing the sampling pattern~\cite{18Landon,16Chao}. Nevertheless in the benchmark experimental results below, we sample the state on an uniformly-spaced square grid in order to learn more about the method and obtain direct pictures of the quantum mechanical oscillator states.
\begin{figure}[tb]
\begin{center}
\includegraphics[width=0.975\columnwidth]{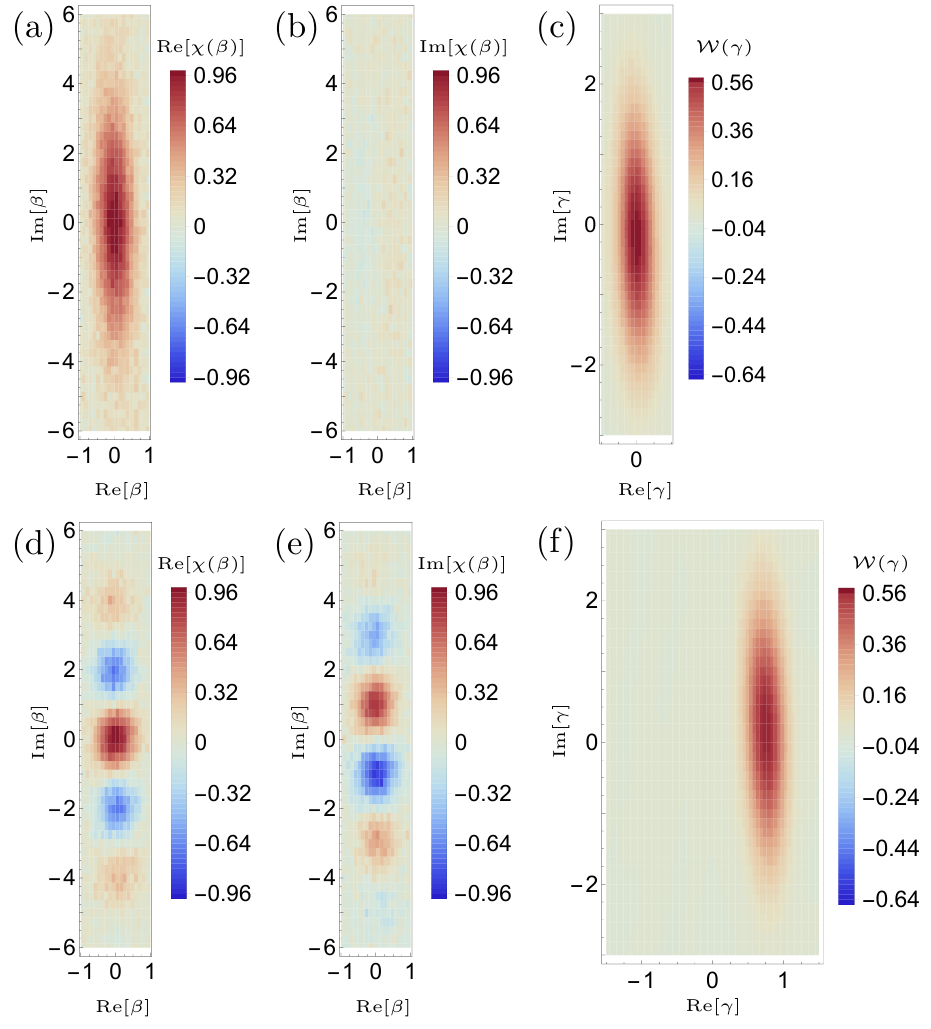}
\end{center}
\caption{Reconstruction of displaced squeezed oscillator states $\ket{\delta,r e^{i\vartheta}}=\Dis{\delta}\Sq{r e^{i\vartheta}}\ket{0}$ with $r_{c}=0.93 \pm 0.02$ and orientations chosen to be $\vartheta_{c}=0$. (a)-(c) shows the squeezed vacuum state with $\delta_c=0$, while (d)-(f) reconstruct the displaced squeezed state with $\delta_{c} = 0.78 \pm 0.05$. The real and imaginary readouts of the characteristic function (a) and (b) ((d) and (e)) show direct fluorescence readout data, while the Wigner function (c) was obtained performing the two dimensional Fourier transform given in equation~\ref{eq:WigFT}.
}
\label{fig:Gaussianstate}
\end{figure}

We start our study by analyzing displaced and squeezed vacuum states $\inoket=\ket{\delta,r e^{i\vartheta}}=\Dis{\delta}\Sq{r e^{i\vartheta}}\ket{0}$. Here the phase-space squeezing operator is defined as $\Sq{\xi=r e^{i\vartheta}}=\exp\left({(-\xi\hat{a}^{\dagger 2}+\xi^*\hat{a}^2)/2}\right)$
and $\ket{0}$ denotes the oscillator ground state. Displaced squeezed states are prepared experimentally by first cooling the ion's motion into a squeezed state $\ket{r e^{i\vartheta}}=\Sq{r e^{i\vartheta}}\ket{0}$ using reservoir engineering~\cite{14Kienzler} and subsequently applying an oscillating voltage to one of our ion trapping electrodes resonant with the ion's motional frequency, realizing a classical force implementing the shift $\Dis{\delta}$. Figure~\ref{fig:Gaussianstate} shows the extracted characteristic function obtained from two states with squeezing parameters $r_{c}=0.93 \pm 0.02$ and $\vartheta_{c}=0$ and displacements $\delta_c = 0$ ((a) and (b)) and $\delta_c = 0.78\pm0.05$ ((d) and (e))  . Here and elsewhere in this manuscript quoted values with subscript $c$ were either set to this value (no error-bar) or are values obtained from independent calibration measurements.

The form of the measured results qualitatively follows the expectation from theory. For tomography in general, a common approach is then to find the physically constrained characteristic function which is closest to the measurement \cite{04Hradil, Dangniam2015}. However identifying a suitable basis set is non-trivial. One assumption could be that states are bounded in energy. A polynomial expansion of the characteristic function would then involve polynomials of order $2m$ for Fock states occupations up to $m$ \cite{Ryl2017}. This approach is complex, and appears rather indirect for obtaining information of experimental interest. As an alternative, we instead look for the closest pure state which might reproduce the data, using a model which takes account of known sources of imperfection, including calibration errors and the presence of state-preparation and measurement errors of the spin state (SPAM).  We fit the measurement data to the functional form
\begin{equation}
E(\beta)= \charfun(\beta)(1-|b|)+b
\label{eq:Efit}
\end{equation}
with $\charfun(\beta) $ a function for the characteristic function based on a small set of parameters $\left\{\xi \right\}$, and $b$ a bias parameter which accounts for (SPAM)~\cite{Supp}. In each case $b$ and  $\left\{\xi \right\}$ are floated. For pure displaced-squeezed states the characteristic function is
\begin{align}
\charfun(\beta,\ket{\delta,r e^{i\vartheta}})=e^{-|\beta\cosh(r)+\beta^*e^{i\vartheta}\sinh(r)|^2/2}e^{\beta\delta^*-\beta^*\delta}.
\label{calcsq}
\end{align}
and $\{\xi\} = \{\delta, r, \vartheta\}$.

We rate the quality of the fit based on a standard reduced chi-squared function
$
c_r=1/(N - \nu)\sum_{i=1}^N (\charfun(\beta_i)-E(\beta_i))^2/\sigma_i^2
$
where $\nu$ denotes the number of fitting parameters, $N$ the total number of measurements, $\charfun(\beta_i)$ the measurement result at the phase-space point $\beta_i$ and $\sigma_i$ the standard error on the mean (s.e.m.) of each point. For the (displaced) squeezed vacuum states the fitted parameters yield ($c_r=1.07$) $c_r = 1.09$, which is a significant improvement over the values  ($c_r=1.78$), $c_r=1.82$ obtained using the independently calibrated values. For both states the fit revealed a small tilt $\vartheta = \Uni{0.044 \pm 0.002}{}$ together with a discrepancy in the shift $|\textrm{Im}[\delta]| = 0.149 \pm 0.006$. In addition a bias $b = \Uni{3.05 \pm 0.07}{\%}$ was found, which was explained due to poorly calibrated internal-state preparation for this data set. Quoted values above denote the average of the fitted parameters for the two states. A complete list of all parameters can be found in table~\ref{tab:FittetPar} of the supplemental material~\cite{Supp}. The tilt $\vartheta$ is only visible in the large data set and indicates the potential for improving the SDF and squeezed state phase calibration in the future. The cause for the small shift along the imaginary axis is currently unclear. Such a shift could arise due to a weak carrier drive during the squeezed state preparation~\cite{Supp}.\par

To obtain a Wigner function for our states, we perform the discrete version of the Fourier transform (DFT) given in equation~\ref{eq:WigFT} with $l=0$. Prior to the DFT, we first subtract the bias $b$ and zero-pad the data outside the measurement range, and additionally re-sample the data on an equidistant grid. Results are shown in part (c) and (f) of figure~\ref{fig:Gaussianstate} respectively. The numerical errors occurring due to the additional data processing can be estimated using sampling of ideal states (see~\cite{Supp} for more details). The average magnitude of the discrepancy over all sampled points of the Wigner function is found for the states above to be $0.29\%$. Comparison of the Wigner and characteristic functions show for both states that these quantities exhibit a smaller extent and hence a reduced uncertainty along $\textrm{Re}(\gamma)$ versus $\textrm{Im}(\gamma)$. However the displacement $\Dis{\delta}$ has different effects, shifting the Wigner function while appearing in the characteristic function as an oscillation of the function in the direction perpendicular to the shift. The latter is due to the geometric phases $\Dis{\beta}\Dis{\delta}=\exp\left({(\beta \delta^*-\beta^*\delta)/2}\right)\Dis{\delta+\beta}$
which occur when displacement operators are combined.

%

\begin{figure}[tb]
\begin{center}
\includegraphics[width=0.975\columnwidth]{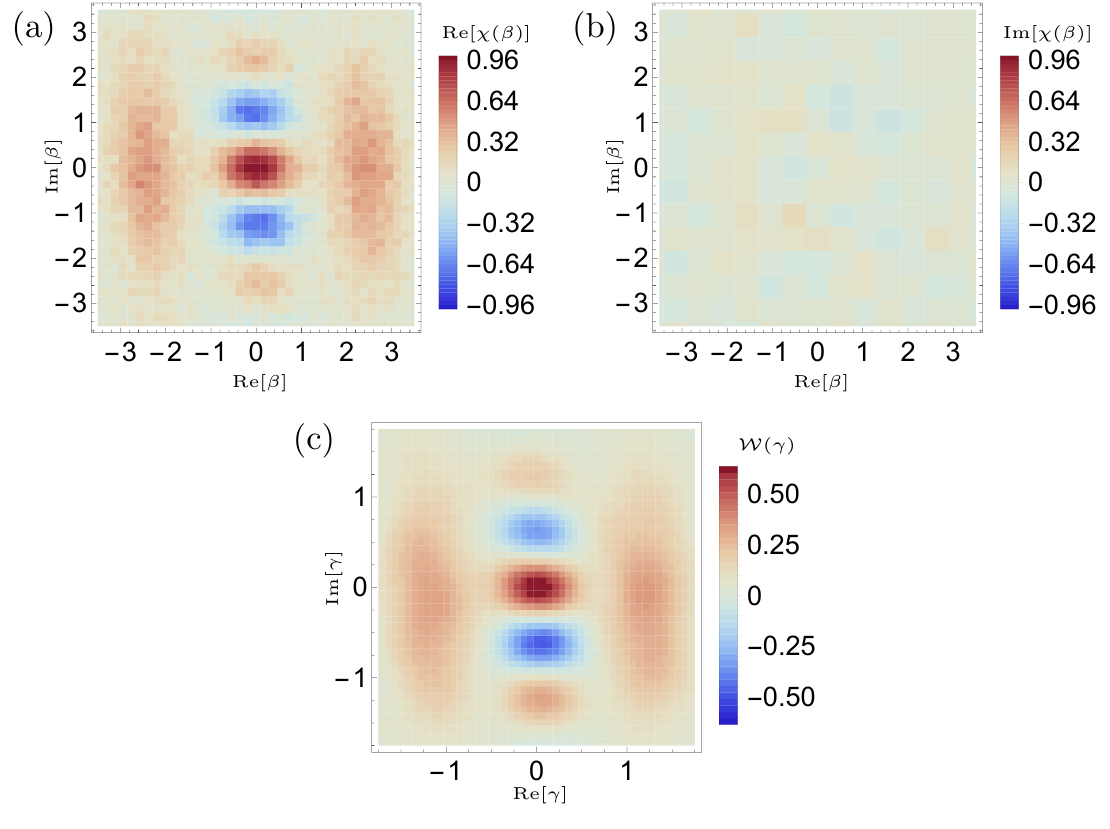}
\end{center}
\caption{Squeezed cat state reconstruction $[\Dis{-\alpha/2}+\Dis{\alpha/2}]\ket{re^{i\vartheta}}$ with $\alpha_c=2.42\pm0.01$, $r_c=0.58\pm0.02$. Calibration and measurement data agree with a reduced chi square of $c_r=2.05$. The similarity between the characteristic function (a) and the Wigner function found via DFT (c) is misleading. The stable phase relation between the two shifted squeezed states is in (a) confirmed by the lobes at $\textrm{Re}[\beta]=\pm 2.42$, while in (c) the stable phase is confirmed by the oscillations along $\textrm{Im}[\beta]$.
}
\label{fig:Cat}
\end{figure}

Displaced and squeezed states belong to the category of Gaussian oscillator states~\cite{SCHUMAKER1986, 05Braunstein}. Non-Gaussian states can exhibit negative points in the Wigner representation, which make interpretations of the Wigner function as a probability density in phase-space impossible. One example of a class of states exhibiting negative Wigner function points are the so called ``cat states'' relating to Schr{\"o}dinger's infamous ``dead and alive cat'' thought experiment~\cite{Schrodinger1935}. Such states can be realized as a superposition of two displaced squeezed states
\begin{align}
\inoket\propto\Dis{\delta}[\Dis{-\alpha/2}+\Dis{\alpha/2}]\ket{re^{i\vartheta}} \ ,
\end{align}
 which are 
created in our experiments by applying a post-selected modular measurement with a squeezed oscillator state as input $\ket{\psi_\textrm{in}}$ \cite{16Kienzler, 15Lo}. Experimental reconstruction of the characteristic function of such a superposition state with $\alpha_c=2.42 \pm 0.01$, $r_c=0.58 \pm 0.02$, $\vartheta_c=\delta_c=0$ is shown in figures~\ref{fig:Cat}. The time scale of this measurement was $\approx \Uni{6}{h}$ which required repeated recalibration of the coupling strength of the bi-chromatic laser pulse. Both the duration of the internal-state dependent shift used for the preparation of the cat state as well as during the analysis were updated accordingly. This leads to additional fluctuations on the input state preparation due to the calibration accuracy, when compared to shorter experiments. We again fit the measurement data to the expected analytic functional form~\cite{Supp}, including the bias $b$, and obtain a reduced chi-squared of $c_r=1.40$ which is a reduction relative to that of the calibrated values $c_r=1.71$. The fit parameters $\textrm{Re}[\alpha] = 2.396 \pm 0.004$, $r = 0.543 \pm 0.005$ are quite close to the calibrated values. A substantially smaller shift $\Uni{|\delta| = 0.04 \pm 0.01}{} $ and bias $b=\Uni{0.9 \pm 0.1}{\%}$ was obtained in this case compared to the displaced-squeezed states. However the tilt $\Uni{\vartheta = 0.110 \pm 0.007}{}$ increased (see~\cite{Supp} for full list of fit parameters). \par

The intrinsic quantum mechanical feature of the cat state is given by the stable phase relation between the differently displaced parts. For the ideal calibrated state with $\delta_c = 0$ this is indicated by the value of the parity $\expect{\Parity}= 1$ where $\Parity\equiv(-1)^{\create \destroy}$~\cite{02Barnett}. This is closely related to the value of the Wigner function at the origin ($\gamma = 0$ in equation \ref{eq:WigFT}) and thus to the integral of the characteristic function over the full space
\begin{align}
\expect{\Parity}=\frac{\pi}{2}\wig(0)=\frac{1}{2\pi}\int  \textrm{Re}[\charfun (\beta)] d^2 \beta.
\label{eq:Parity}
\end{align}
Performing the Fourier transform of the characteristic function measurements as above gives a parity of $\expect{\Parity} = 0.98$ (we find the numerical-analysis DFT error in this case to be $0.70\%$, see \cite{Supp}). In this estimation the bias due to SPAM plays an important role. Without subtracting the fitted value of $b = \Uni{0.9\pm0.1}{\%}$ from all data we find a value of $\expect{\Parity} = 0.90$. This example shows that any constant offset in the data leads to large error in a measurement of the parity. It is worth noting that the close similarity between the characteristic function and the Wigner function for the squeezed-cat state superposition is misleading. The oscillation in the Wigner function along the imaginary axis indicate the presence of a stable phase relation between the two displaced components. In contrast the oscillations along the imaginary axis in the characteristic function would be identical for the mixture $\rho_\textrm{mix}\propto\ketbra{\alpha/2,\xi}{\alpha/2,\xi}+\ketbra{-\alpha/2,\xi}{-\alpha/2,\xi}$. In case of the characteristic function the phase relation is confirmed by the peaks at $\textrm{Re}[\beta] = \pm \alpha$.\par

\begin{figure}[tb]
\begin{center}
\includegraphics[width=0.975\columnwidth]{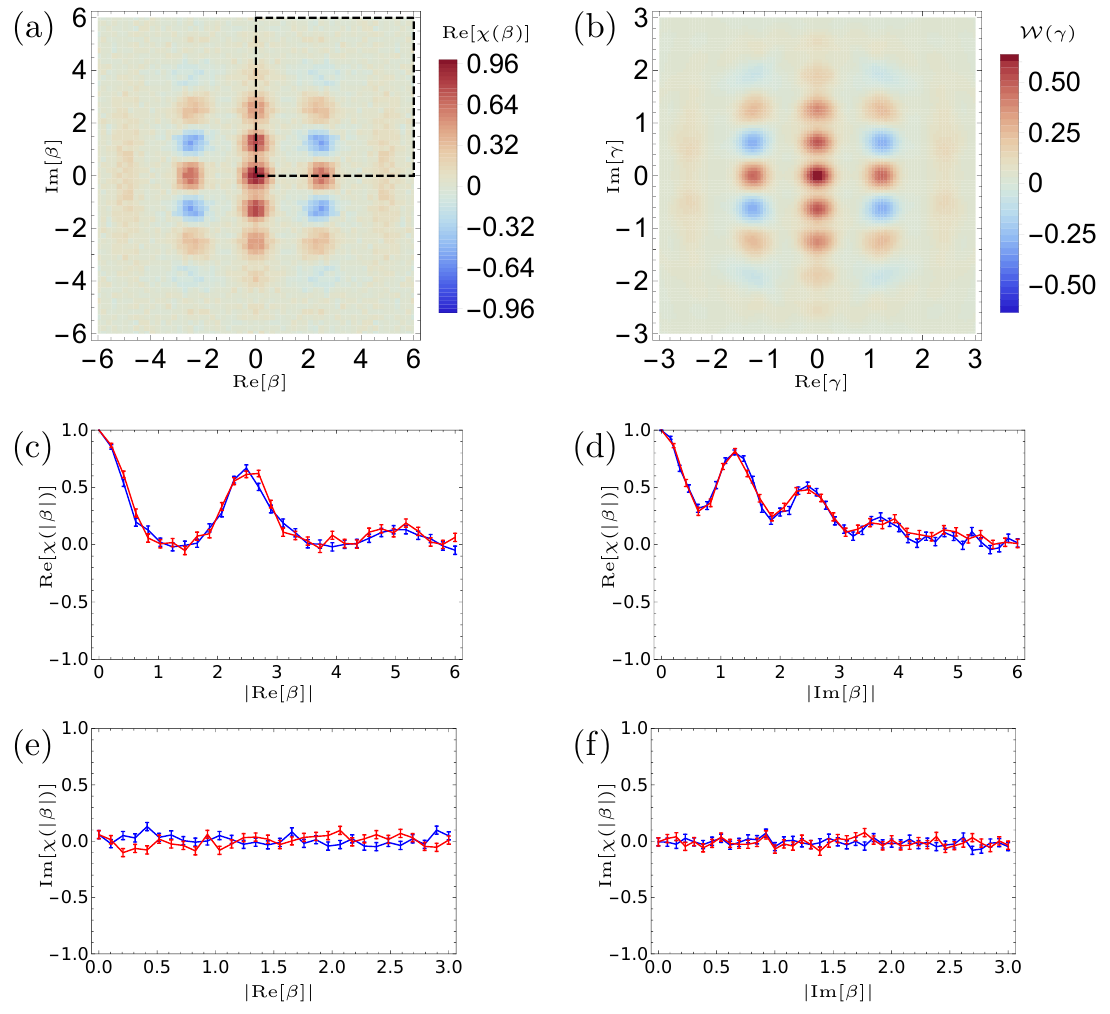}
\end{center}
\caption{Reconstruction of a three component superposition state $\Dis{\delta}[\Dis{-l}+2\cdot\mathbb{1}+ \Dis{l}]\ket{re^{i\vartheta}}$ previously used to encode a qubit in the trapped-ions motion. The state was calibrated to $l_{\textrm{c}}=2.50 \pm 0.05$ and $r_{\textrm{c}}=0.93\pm 0.03$, $\delta_{\textrm{c}}=\vartheta_{\textrm{c}}=0$ calibration and measurements agree with $c_r=2.05$. In order to reduce experimental run time only the positive quadrant of $\textrm{Re}[\charfun (\beta)]$ was measured of the symmetric state. The measured area is indicated by the dashed box in (a) and was combined with three mirrored versions of itself to visualize the full state. (b) Shows the Wigner function obtained via Fourier transform. In (c)-(f) we confirm the symmetries of the state with scans along the complete axes $\textrm{Re}[\beta]$ and $\textrm{Im}[\beta]$. Errors are given as standard errors of the mean. We overlay results of the negative axis (blue) with results of the positive axis (red) by plotting measurements as a function of $|\beta|$. The measurements of $\textrm{Re}[\charfun (\beta)]$ (c) and (d) exhibit even symmetrie within the s.e.m errors while $\textrm{Im}[\charfun (\beta)]$ measurements (e) and (f) are close to zero.
}
\label{fig:Grid}
\end{figure}

After testing the characteristic function method we tackle partial reconstruction of a three-component superposition $\inoket\propto\Dis{\delta}[\Dis{-l}+2\cdot\mathbb{1}+ \Dis{l}]\ket{re^{i\vartheta}}$ with $l_{c}=2.50 \pm 0.05$ and $r_{c}=0.93\pm 0.03$, $\delta_{c}=\vartheta_{c}=0$, which is an approximate GKP code state \cite{18Fluhmann2, Devoret}. Figure~\ref{fig:Grid} shows the characteristic function measurement results. In this case only the positive quadrant of the real part was measured, which is indicated by the dashed box in figure \ref{fig:Grid} (a). The complete plot is obtained combining the measured quadrant with three mirrored copies. From these measurements the parity is estimated to be $\expect{\Parity} \approx 0.95\pm 0.02$ with the error bar denoting the error due to the uncertainty in $b$. This is close to 1 and thus constrains the value of the imaginary part to be close to zero throughout. The expected symmetries of the state are additionally confirmed by measuring the full imaginary and real phase space axes. Results are shown in (c) - (f) respectively, where the negative axis (blue) is overlaid with the positive axis (red). The measurements confirm that the essential information about the reconstructed state is captured by the positive quadrant. Part (e) however shows a small systematic close to zero, which only partially follows the expected odd symmetry of $\textrm{Im}[\charfun]$. This might be due to small uncontrolled and partially fluctuating displacements of the state from the origin $\delta$. Figure~\ref{fig:Grid} (b) shows the respective Wigner function obtained using a DFT. In this case we obtain an average expected error from the numerical analysis of $0.45\%$. In case, a fit using equation \ref{eq:Efit} obtains a reduction of the reduced chi square to $c_r=1.58$ compared to $c_r=2.05$ for the calibrated values. The primary discrepancy between calibration and fit is again a tilt of the squeezing direction $\Uni{\vartheta=0.103\pm0.008}{}$. All other parameters are within the calibration error bars \cite{Supp}.\par

We have implemented direct, simple and versatile reconstruction of the symmetric characteristic function and demonstrated its capabilities by analyzing displaced squeezed states and superpositions of these states. We focused our discussion on potential improvements of experimental control based on the large reconstruction data set revealing discrepancies between calibrations and fits. However it is also worth pointing out that these discrepancies are small: the square fidelities between calibrated and fitted pure states are above $0.98$ for all four analyzed states including the GKP code word \cite{Supp}. Possible extensions of this work include extracting symmetrically ordered expectation of powers of creation and destruction operators~\cite{02Barnett}
\begin{align}
\expect{\create^m\destroy^n}_S=\int_{-\infty}^{\infty} \textrm{d} \gamma^2 \wig(\gamma)\gamma^{*m}\gamma^n\\
\expect{\create^m\destroy^n}_S=\left(\frac{\partial}{\partial \beta}\right)^m \left(-\frac{\partial}{\partial \beta^*}\right)^n \chi(\beta)|_{\beta=0}\notag
\end{align}
which would allow the values of eg. $\nbar$ or $\textrm{g}^2(0)$ of the states to be determined. This would require numerical techniques optimized for  estimating  derivatives from a sparsely sampled noisy signal, which requires filtering of the projection noise in an appropriate way~\cite{Cullum71, Anderssen1974, OppenheimAlanV2014Sas}. Other extensions include the use of optimized sampling patterns, the use of feedback to improve the quality of state preparation. The basic idea of utilizing qubit state-dependent shifts for reconstruction of a bosonic degree of freedom is applicable to a wide variety of spin-boson systems, including various mechanical systems and microwave or optical cavities coupled to superconducting or atomic qubits.

We thank M.~Marinelli, V.~Negnevitsky and T.~Behrle for contributions to the apparatus, and T-L.~Nguyen for useful discussions. We acknowledge support from the Swiss National Science Foundation through the National Centre of Compeentce in Research for Quantum Science and Technology (QSIT) grant 51NF40--160591, and from the Swiss National Science Foundation under grant number $200020\_165555/1$.


\iftoggle{arXiv}{
}{\bibliography{./myrefsStateTomo}}
\iftoggle{arXiv}{

\iftoggle{arXiv}{
	\clearpage
	{\LARGE\bfseries Supplementary Information\\}
}{}

\section{Q function of cat state}
We consider for the cat state a superposition of the form
\begin{align}
\ket{\psi_{\varphi}} =\frac{1}{\sqrt{N}}(\ket{0}+e^{i\varphi}\ket{\alpha})
\label{osc_cat}
\end{align}
with the normalization $N=2(1+\cos(\varphi)e^{-|\alpha|^2/2})$. 

The \qfun function for such a superposition is
\be
\qfun (\beta) &=&\frac{1}{\pi N}\left[e^{-|\beta|^2}+e^{-|\beta-\alpha|^2}\right.\nonumber \\ &+& \left. e^{-|\beta|^2-|\alpha|^2/2}(e^{-i\varphi+\alpha^*\beta}+e^{i\varphi+\alpha\beta^*})\right]  \ .
\label{ho_qcat}
\ee
The  Wigner function for the same state is
\be
\wig (\gamma) &=&\frac{2}{\pi N}\left[e^{-2|\gamma|^2}+e^{-2|\alpha-\gamma|^2} \right. \nonumber \\&+& \left. e^{-2|\gamma|^2-|\alpha|^2/2}(e^{-i\varphi+2\alpha^*\gamma}+e^{i\varphi+2\alpha\gamma^*})\right]
\label{ho_wigcat}
\ee
For simplicity we consider cases where $\alpha \in \mathbb{R}$. The oscillations of the Wigner function positioned midway between the separated wave packets are found centred around  $\gamma = \alpha/2$ and the oscillations are aligned with the imaginary axis. Mathematically these are due to the third term in the equation above. If the imaginary component of the phase space position is given by the parameter $m$, then considering only positions $\gamma = \alpha/2 + i m$ we find
\begin{align}
\wig (\alpha/2 +i m) =& \frac{4}{\pi N}e^{-2|m|^2}[e^{-|\alpha|^2/2}+\cos(\varphi+2m\alpha)]\\
\approx&\frac{4}{\pi N}e^{-2|m|^2}\cos(\varphi+2m\alpha)
\label{ho_wigcat}
\end{align}
with the approximation valid  $\alpha \gg 1$. For the $\qfun$ function we find:
\begin{align}
\qfun (\alpha/2 +i m) =& \frac{2}{\pi N}e^{-|\alpha|^2/4-|m|^2}[1+\cos(\varphi+m\alpha)]\\
\approx & 0
\label{ho_qcat}
\end{align}
in the same limit. Thus we see that the $\qfun$ function also exhibits oscillations along the imaginary axis at $\beta=\alpha/2+i m$, which are only present in case of stable quantum superposition phase $\varphi$. However in contrast to the Wigner function the oscillations are suppressed by the exponential factor $e^{-|\alpha|^2/4}$ which for large separations is close to zero. Thus for large cat states, in the presence of experimental noise and finite accuracy measurement, the information about the coherent nature of the superposition gets lost in the $\qfun$ function representation.

\vspace{10mm}
\section{Displaced Fock based reconstruction}
\label{dispFockrec}
In earlier experiments we performed Wigner function tomography based on the extraction of displaced Fock state populations $\pop(n_\gamma)=|\bra{n}\Dis{-\gamma}\oscin|^2$~\cite{16Kienzler}. The Wigner function is obtained from these Fock state populations as~\cite{96Leibfried}:
\begin{align}
\wig(\gamma) = 2/\pi  \sum_n (-1)^n \pop(n_\gamma).
\end{align}
An appropriately chosen laser pulse (a combination of carrier and blue-sideband resonant drives ~\cite{16Kienzler}) applied to the ion prepared in $\ket{\uparrow}\oscin$ leads to oscillations in the internal state probabilities according to:
\begin{align}
[P(\uparrow, t)-P(\downarrow, t)]=\sum_{n=0}^{\infty} \pop(n_{\gamma}) \cos(\Omega \sqrt{n_{\gamma}+1} t)
\end{align}
with $\Omega/(2\pi)\approx\Uni{20}{kHz}$. Fitting of the measured traces to this functional form floating the populations $\pop(n_{\gamma})$ thus allows to estimate the Wigner function. 
A simple estimate how long an analogous reconstruction to the one presented in figure~\ref{fig:Grid} would take using the old method, leads to at least a factor $R \approx 20 \approx T_{\textrm{wig}}/(p_{ps} \cdot T_{\textrm{char}})$  longer. This estimate is based on using the duration $T_{\textrm{wig}}$ required for reconstruction of one point in~\cite{16Kienzler} for the result shown in figure 4 (a) of a previous publication in which we performed tomography of a cat states ~\cite{16Kienzler}, and comparing this to the duration of reconstruction of one point for the GKP state. The GKP state requires two post-selected modular measurements, which is one more than the cat state. This is factored out in our comparison by taking account of the  probability for successful post-selection in the second measurement $p_{ps}\approx 0.75$. \par
The measurements presented in the current manuscript also include phase space points of larger distance from the origin than previously realized. These become increasingly challenging using the Fock-state reconstruction method since $\pop(n_{\gamma})$ contains high excitations, which lead to increasingly fast oscillations in the internal state populations, while adjacent Rabi frequencies also become closer together in fractional frequency as $n$ increases (this scales as $1/(2 n)$ for $n \gg 1$). Figure \ref{figSupp} shows $\pop(n_{\gamma})$ for a GKP state of the size considered in the current manuscript, sampled at displacements of $\gamma=0$, $\gamma=1.5+1.5i$ and $\gamma=3+3i$, which range up to $n \simeq 50$. For comparison fits in reconstruction of figure 4 (a) of~\cite{16Kienzler} included populations up to 35.\par
Previously for cat states we had managed to reconstruct cat states with large displacements by measuring in a squeezed Fock basis with the anti-squeezing axis aligned with the cat displacement ~\cite{16Kienzler}. This trick is however not straightforward to apply to the GKP states, which have significant extent in two perpendicular directions in phase space. 
\begin{figure*}[tb]
\begin{center}
\includegraphics[width=1.03\textwidth]{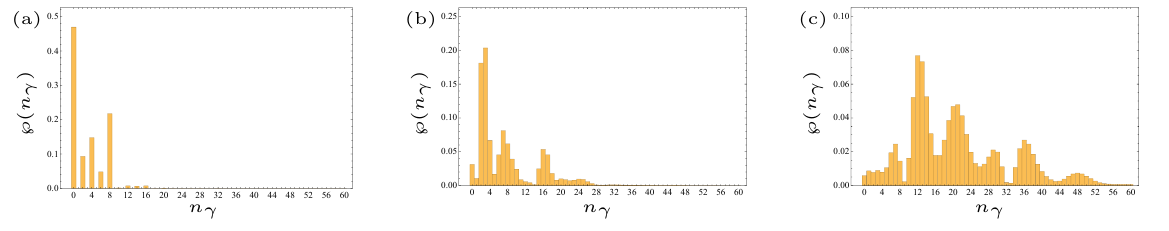}
\end{center}
\caption{Displaced Fock state populations of three component GKP state $\pop(n_{\gamma})=|\bra{n}\Dis{-\gamma}\oscin|^2$. (a) $\gamma=0$, (b) $\gamma=1.5+1.5i$, (c) $\gamma=3+3i.$
}
\label{figSupp}
\end{figure*}

\section{Calibration routines}
 The calibration and state preparation routines used in this work have been used previously, and are described in more detail in the publications and appendices of~\cite{18Fluhmann2,18Fluhmann,16Kienzler,15Lo,14Kienzler}. Below I provide a short summary for the techniques relevant to this work.

\subsection{State-dependent force and separation amounts $\alpha_c$ and $l_c$}
The internal state-dependent force realizing the operations $\Dis{\alpha(t)\sigx}$, where $\alpha(t)=\eta\Omega t e^{-i\Delta\phi/2}$ is realized by driving the blue and red sideband of the internal state transition simultaneously with equal strength. This leads to the Hamiltonian $\hat{H}_{\text{SDF}}=\eta \hbar \Omega \hat{ \sigma}_x(\hat a e^{i\Delta\phi/2}+\hat{ a}^{\dagger}e^{-i\Delta\phi/2})/2$. The resulting shift in phase space is proportional to the Rabi frequency and the duration of the pulse $\alpha(t)\propto \Omega t$. We calibrate the proportionality between the shift and the pulse duration by comparing the induced shift to known oscillator states of fixed extent. Typically we use Fock state $\ket{n=1}$, which we create by ground-state cooling followed by a blue sideband $\pi$-pulse transforming $\ket{\downarrow}\ket{0}$ to $\ket{\uparrow}\ket{1}$. We then subsequently apply the SDF pulse for variable duration $t$. We fit the recorded internal state fluorescence to its expected form $\expect{\hat{Z}(t)}=1-e^{-2(c t)^2}(1-(2ct)^2)$ where we float the proportionality constant $c$. Typical values of $c\approx\Uni{36 \pm 0.4}{ms^{-1}}$ are obtained. $c$ is then used as a proxy for the Rabi frequency $\Omega$, such that $\alpha(t) = c t$.

\subsection{Squeezing $r_c$ and orientation $\vartheta_c$}
After pumping the ions motion into a squeezed state~\cite{14Kienzler} and preparing the internal state in $\intup$ ($\intd$) we apply a red sideband (blue sideband) of variable duration $t$ and record the oscillations of the internal state-populations. As described in the previous section~\ref{dispFockrec} and in~\cite{96Leibfried,16Kienzler}
this allows to extract the Fock state populations of the oscillator state. We fit the populations to the expectation for a squeezed state while floating the squeezing amplitude $r_c$.\par
This extraction of the Fock state populations is independent of the squeezed state orientation $\vartheta_c$. Experimentally this orientation is controlled via the difference phase of blue and red sideband laser phases during the reservoir engineering. These are the same parameters which define the direction $\arg(\alpha)$ of the SDF. Previous comparisons have shown no measurable deviation from the expected relative orientation. Thus we simply set $\vartheta_c$ to the desired value. The larger data sets collected in this work show small deviation between SDF laser pulse and squeezed state orientation. By choosing appropriate settings, this observation might allow to improve calibrations in the future.

\subsection{Phase space shift $\Dis{\delta}$ calibration}
Internal state independent shifts $\Dis{\delta}$ are realized by a classical electric force to a trapping electrode, the signal for which is generated in the same multi-channel radio-frequency source unit as the SDF. To account for mismatches in the signal paths the orientation of the shift has to be calibrated relative to the SDF. Combining the state-dependent force with internal state rotations allows to realize an effective state independent shift using the laser. For the calibration of the electric shift to the laser shift we typically use a squeezed probe state. We first displace this using the SDF by a calibrated amount, and then calibrate the electric force in order to invert the displacement. See Methods of~\cite{18Fluhmann2} for more details.

\subsection{Motional frequency calibration}
Key to the presented experiments is a stable and well calibrated laser frequency relative to the motional oscillation frequency. We typically recalibrate the motional frequency every \Uni{5}{min} for \Uni{20}{s} achieving an accuracy of around $\Uni{\pm 20}{Hz}$. The calibration sequence in this case is given by ground state cooling followed by a displacement $\Dis{\beta}$, a long wait time $T$ and a second displacement $\Dis{-\beta}$. The coherent state only returns to the ground state in case of a well calibrated motional frequency. Using a red sideband we probe for which frequency the state returns to the origin.  The calibration technique is described in detail in the Methods of~\cite{18Fluhmann2}.

\section{Analytic forms for ideal states}
{
\begin{table*}[h]
\begin{tabular}{|l|r@{\;{=}\;}r@{\;{$\pm$}\;}l|r@{\;{=}\;}l|r@{\;{=}\;}r@{\;{$\pm$}\;}l|r@{\;{=}\;}l|l|l|}
\hline
\multicolumn{1}{|c|}{state}	 &\multicolumn{3}{c|}{calibrated $\ket{\psi}_c$}	&\multicolumn{2}{c|}{red. chi}	&\multicolumn{3}{c|}{fitted $\ket{\psi}$} &\multicolumn{2}{c|}{red. chi} & \multicolumn{1}{c|}{dur.}	&\multicolumn{1}{c|}{fid.} \\\hline
squeezed state									& $r_c$ & $0.93$ & $0.02$ 			& $c_r$ & $1.82$ 					& $r$ & $0.938$ &$0.005$ 			 & $c_r$ & $1.09$ 					& 1\: h&0.993\\
$\ket{\delta,re^{i\vartheta}}$	& $\vartheta_c$ & \multicolumn{2}{l|}{0} 	&	\multicolumn{2}{c|}{}		&	$\vartheta$ & $0.041$ & $0.003$ & \multicolumn{2}{c|}{} 	&&\\
																					& $\delta_c$			& \multicolumn{2}{l|}{0}				&	\multicolumn{2}{c|}{}	& $\textrm{Re}[\delta]$ & $0.003$ & $0.001$ & \multicolumn{2}{c|}{} &&\\
																					& \multicolumn{3}{c|}{}				&	\multicolumn{2}{c|}{}		& $\textrm{Im}[\delta]$ & $-0.184$ & $0.009$ & \multicolumn{2}{c|}{} &&\\
																					& \multicolumn{3}{c|}{}				&	\multicolumn{2}{c|}{}		& $b$ 			& $0.035$ &$0.001$ & \multicolumn{2}{c|}{} &&\\ 	\hline
																		
displaced squeezed state				& $r_c$ & $0.93$ &$0.02$ 			&  $c_r$ & $1.78$ 					& $r$ & $0.925$ & $0.004$ 			& $c_r$ & $1.07$ 				& 1\: h &0.992\\
$\ket{\delta,re^{i\vartheta}}$	& $\vartheta_c$ & \multicolumn{2}{l|}{0}	&	\multicolumn{2}{c|}{}		&	$\vartheta$ & $0.047$ & $0.003$ & \multicolumn{2}{c|}{} 	&&\\
																					& $\textrm{Re}[\delta_c]$			& $0.78$	& $0.05$				&	\multicolumn{2}{c|}{}		& $\textrm{Re}[\delta]$ & $0.752$ & $0.001$ & \multicolumn{2}{c|}{} &&\\
																					& $\textrm{Im}[\delta_c]$			& $0$	& $0.07$								&	\multicolumn{2}{c|}{}		& $\textrm{Im}[\delta]$ & $0.114$ & $0.008$ & \multicolumn{2}{c|}{} &&\\
																					& \multicolumn{3}{c|}{}				&	\multicolumn{2}{c|}{}		& $b$ 			& $0.026$ &$0.001$ & \multicolumn{2}{c|}{} &&\\\hline
																					
squeezed cat	& $r_c$ & $0.58$ &$0.02$ 		&  $c_r$ & $1.71$ 				& $r$ & $0.543$ & $0.005$ 			& $c_r$ & $1.40$ 				& 6\: h& 0.989\\
$\Dis{\delta}[\Dis{\alpha/2}+\Dis{-\alpha/2}]\ket{re^{i\vartheta}}$& $\vartheta_c$ & \multicolumn{2}{l|}{0} 		&	\multicolumn{2}{c|}{}																&	$\vartheta$ & $0.110$ & $0.007$ & \multicolumn{2}{c|}{} 	&&\\
																					& $\alpha_c$			& $2.42$	& $0.01$		&	\multicolumn{2}{c|}{}												& $\textrm{Re}[\alpha]$ 		& $2.398$ & $0.004$ & \multicolumn{2}{c|}{} &&\\
																					& $\delta_c$			& \multicolumn{2}{l|}{0}						&	\multicolumn{2}{c|}{}												& $\textrm{Im}[\alpha]$ 		& $-0.009$ & $0.012$ 		& \multicolumn{2}{c|}{} &&\\
																					& \multicolumn{3}{c|}{}									&	\multicolumn{2}{c|}{}												& $\textrm{Re}[\delta]$ 		& $0.020$ 		& $0.006$ & \multicolumn{2}{c|}{} &&\\
																					& \multicolumn{3}{c|}{}									&	\multicolumn{2}{c|}{}												& $\textrm{Im}[\delta]$ 		& $-0.031$ 		& $0.007$ & \multicolumn{2}{c|}{} &&\\
																					& \multicolumn{3}{c|}{}									&	\multicolumn{2}{c|}{}												& $b$ 			& $0.009$ &$0.001$ & \multicolumn{2}{c|}{} &&\\\hline
																					
GKP state																	& $r_c$ & $0.93$ &$0.03$ 		&  $c_r$ & $2.05$ 				& $r$ & $0.892$ & $0.008$ 			& $c_r$ & $1.58$ 				& 6\: h& 0.985\\
$\Dis{\delta}[\Dis{-l}+2\cdot\mathbb{1}+ \Dis{l}]\ket{re^{i\vartheta}}$	& $\vartheta_c$ & \multicolumn{2}{l|}{0} 		&	\multicolumn{2}{c|}{}																&	$\vartheta$ & $0.103$ & $0.008$ & \multicolumn{2}{c|}{} 	&&\\
																					& $l_c$			& $2.50$	& $0.05$		&	\multicolumn{2}{c|}{}																& $\textrm{Re}[l]$ 		& $2.471$ & $0.005$ & \multicolumn{2}{c|}{} &&\\
																					& $\delta_c$			& \multicolumn{2}{l|}{0}	&	\multicolumn{2}{c|}{}																& $\textrm{Im}[l]$ 		& $0.022$ & $0.008$ & \multicolumn{2}{c|}{} &&\\
																					& \multicolumn{3}{c|}{}				&	\multicolumn{2}{c|}{}																& $\textrm{Re}[\delta]$ 		& $0.001$ 		& $0.003$ 		& \multicolumn{2}{c|}{} &&\\
																					& \multicolumn{3}{c|}{}				&	\multicolumn{2}{c|}{}																& $\textrm{Im}[\delta]$ 		& $-0.002$ 		& $0.007$ 		& \multicolumn{2}{c|}{} &&\\
																					& \multicolumn{3}{c|}{}				&	\multicolumn{2}{c|}{}												& $b$ 			& $0.0015$ &$0.001$ & \multicolumn{2}{c|}{} &&\\\hline

\end{tabular}
\caption{Reconstructed states calibrated and fitted parameters. The first column denotes the type of state we aimed to reconstruct. In the second column the independently calibrated parameters are summarized and the third column compares the calibration to the reconstruction by calculating the reduced chi square $c_r$. The fourth column provides the fit parameter results. In the fifth column the reduced chi square between the fitted state and the measurement data is assessed. The sixth column shows approximate measurement times, while in the last column the state fidelity between the calibrated and fitted pure states is calculated $|\bra{\psi}_c\ket{\psi}|^2$.}
\label{tab:FittetPar}
\end{table*}
}

For each of the states created in our experiments, both the characteristic and the Wigner function can be calculated analytically. For an arbitrary superposition $\ket{\psi}$ of coherent states $\ket{\delta}=\Dis{\delta}\ket{0}$, we can write
\begin{align}
\genosc&=\frac{1}{\sqrt{N}}\sum_{\delta} c_{\delta}\ket{\delta}\\
N&=\sum_{\delta,\epsilon} c_{\delta}^* c_{\epsilon} e^{-\frac{1}{2}(|\delta|^2+|\epsilon|^2-2\delta^*\epsilon)}
\end{align}
where we obtain the normalization using
\begin{align}
\label{eq:coherent_overlap}
\braket{\beta}{\alpha}=e^{-\frac{1}{2}(|\beta|^2+|\alpha|^2-2\beta^*\alpha)}
\end{align}
The characteristic function then follows as:
\begin{align}
\chi(\beta)&=\bragenosc \Dis{\beta}\genosc =\frac{1}{N}\sum_{\delta,\epsilon} c_{\delta}^* c_{\epsilon} \bra{\delta}\Dis{\beta}\ket{\epsilon}\\
&=\frac{1}{N}\sum_{\delta,\epsilon} c_{\delta}^* c_{\epsilon} e^{-\beta^*\epsilon +\delta^*\beta +\delta^*\epsilon}e^{-\frac{1}{2}(|\delta|^2+|\beta|^2+|\epsilon|^2)}.
\end{align}
and performing the Fourier transform equation~\ref{eq:WigFT} we find the Wigner function:
\begin{align}
\wig(\beta)&=\frac{2}{N\pi}\sum_{\delta,\epsilon} c_{\delta}^* c_{\epsilon} e^{-\delta^*\epsilon+2\delta^*\beta+2\epsilon\beta^*-2|\beta|^2}e^{-\frac{1}{2}(|\delta|^2+|\epsilon|^2)}.
\end{align}
For the superpositions of displaced squeezed states we use the relations~\cite{02Barnett}
\begin{align}
\charfun(\beta,\Sq{\xi}\genosc)&=\charfun(\beta \cosh{(r)}+\beta^*e^{i\vartheta}\sinh{(r)},\genosc)\\
\wig(\gamma,\Sq{\xi}\genosc)&=\wig(\gamma \cosh{(r)}+\gamma^*e^{i\vartheta}\sinh{(r)},\genosc)\\
\end{align}
where we make use of the ability to interchange squeezing and displacement operators using
\begin{align}
\label{change_sqdis}
\Sq{\xi}\Dis{\alpha'}\ket{0}=\Dis{\alpha}\Sq{\xi}\ket{0}\\
\alpha' = \textrm{cosh}(r) \alpha + e^{i\vartheta} \textrm{sinh}(r) \alpha^{\ast}
\end{align}.

\subsection{Calculation fidelities between calibrated and fitted states}
All overlaps given in the last column of table~\ref{tab:FittetPar} can be calculated analytically. One example for how to do this uses the expansion of the states in the Fock state basis:
\begin{align}
\ket{\psi}=\sum_{n=0}^\infty c_n\ket{n}
\end{align}
All four states are given by superpositions of displaced squeezed states $\ket{\beta,\xi}$, using the formula
\begin{widetext}
\begin{align}
\bra{n}\Dis{\beta}\Sq{\xi}\ket{0}&=\exp\left(-\frac{1}{2}|\beta|^2-{\beta^*}^2\frac{e^{i\vartheta}}{2}\tanh(r)\right)i^n\sqrt{\frac{e^{in\vartheta}}{n!\cosh(r)}}\left(-\frac{\tanh(r)}{2}\right)^{n/2}\notag\\
&\times H_n \left[-\frac{i}{2}e^{-i\vartheta/2}\sqrt{\frac{2}{-\tanh(r)}}(\beta+e^{i\vartheta}\tanh(r) \beta^*)\right]
\end{align}
\end{widetext}
which is drawn from equation 2.48 of reference~\cite{agarwal_2012} adapted to our definition of the squeezing operator and corrected for typos. This allows to find the analytic forms for the coefficient $c_n$ for all four states. Using this result, the Fidelity can be calculated as
\begin{align}
F=|\bra{\psi_a} \psi_b\rangle|^2=|\sum_{n=0}^{n_{\textrm{max}}} a_n^*\bra{n}\sum_{m=0}^{m_{\textrm{max}}} b_m\ket{m}|^2=|\sum_{n=0}^{n_{\textrm{max}}} a_n^* b_n|^2 \ .
\end{align}
In our calculations, the truncation of the Fock state expansion is at the value $n_{\textrm{max}}=500$. In order to check our result we calculate the same fidelities based on the formula for $\braket{\beta_1,\xi_1}{\beta_2,\xi_2}$ equation~3.18 from~\cite{Kral90} (which was also corrected for minor typos).

\subsection{Estimation of numerical errors in performing the discrete Fourier transform}
We assess numerical errors due to sampling and DFT by applying our data analysis to numerical data produced by sampling characteristic function values from ideal states with the calibration settings used in the experiment. The analytically calculated ideal characteristic function is sampled at the same phase-space points as the measurement data. We then account for experimental projection noise by drawing a random sample $samp$ from the binomial distribution $B[N,P(1)]/N$, where $N$ is the number of measured fluorescence readout shots and the expected upper state detection probability $P(\uparrow)$ is related to the analytically calculated characteristic function as: $P(\uparrow)=(\charfun+1)/2$. Then we compare the discrete Fourier transform of all sampled characteristic function points ($\charfun=2\cdot samp -1$) to the analytic calculation of the Wigner function of the calibrated state. The average deviation in the presented points in the figures~\ref{fig:Gaussianstate}-\ref{fig:Grid} of the main text expressed in percent with respect to the full Wigner function range $4/\pi$ are: $0.29\%$ for the squeezed vacuum, $0.28\%$ for the displaced squeezed state, $0.70\%$ for the squeezed cat and $0.45\%$ for the GKP state.
Note that full state information is provided by the characteristic function measurements themselves. The discrete Fourier transform is solely performed for illustrative purposes.


\section{State preparation infidelities}
Our experimental system operates with \caf as well as \bef ions. For \bef a first order magnetic field insensitive transition is available at a magnetic field value of $\Uni{115.45}{G}$~\cite{05Langer}. Therefore all experiments, including experiments based only on \caf ions run at this relatively high (compared to typical \caf setups) magnetic field value. The larger splittings of internal state levels (i.e. $\approx\Uni{200}{MHz}$ between neighboring levels in the $D_{5/2}$ manifold or \Uni{335}{MHz} in $S_{1/2}$) pose additional challenges in initial state preparation, repumping and shelving. Squeezed state pumping for example is based on coherent manipulations of the qubit transition together with induced decay of the upper qubit level $\intup (D_{5/2}, m_j=3/2)$. The latter is implemented by driving $\intup$ to the $P_{3/2}$ manifold using a \Uni{854}{nm} laser. The $P_{3/2}$ manifold has a non zero probability to decay back to the $D_{5/2}$ manifold instead $S_{1/2}$. Therefore multiple frequency components and polarizations are used in the \Uni{854}{nm} laser in order to drive all Zeeman sub-levels of $D_{5/2}$ to $P_{3/2}$~\cite{Th:Kienzler}. A weak or poorly calibrated \Uni{854}{nm} laser leaves population behind in the $D_{5/2}$ manifold and can lead to initial state preparation infidelities. These infidelities in particular are also presented after conditioning on a dark detection result in the creation of a superposition state. These considerations motivate our choice of the bias $b$ parameter in our model function.

\iftoggle{arXiv}{
}{\bibliography{./myrefsStateTomo}}
}{}

\end{document}